\documentclass{article}

\begin{document}

\begin{center}
\bigskip How to Complete the Description of Physical Reality by Non-local Hidden Variables?

\textit{Timur F. Kamalov}

Physics Department\\[0pt]
Moscow State Open University\\[0pt]
107996 Moscow, 22, P. Korchagin str., Russia\\[0pt]
E-mail: ykamalov@rambler.ru\\[0pt]
www.TimKamalov.narod.ru\\[0pt]
\end{center}

\begin{abstract}
Which non-local hidden variables could complement the description
of physical reality? The present model of extended Newtonian
dynamics (MEND) is generalize but not alternative to Newtonian
Dynamics because its extended Newtonian Dynamics to arbitrary
reference frames. It Is Physics of Arbitrary Reference Frames.
Generalize and alternative is not the same. MEND describes the
dynamics of mechanical systems for arbitrary reference frames and
not only for inertial reference frames as Newtonian Dynamics.
Newtonian Dynamics can describe non-inertial reference frames as
well introducing fiction forces. In MEND we have fiction forces
naturally and automatically from new axiomatic and we needn't have
inertial reference frame. MEND is differs from Newtonian Dynamics
in the case of micro-objects description.
\end{abstract}


\vspace{5mm}

\section{Introduction}

Classical Newtonian mechanics is essentially the simplest way of
mechanical system description with second-order differential
equations, when higher order time derivatives of coordinates can
be neglected. The extended model of Newtonian mechanics with
higher time derivatives of coordinates is based on generalization
of Newton's classical axiomatics onto arbitrary reference frames
(both inertial and non-inertial ones) with body dynamics being
described with higher order differential equations. Newton's Laws,
constituting, from the mathematical viewpoint, the axiomatics of
classical physics, actually postulate the assertion that the
equations describing the dynamics of bodies in inertial frames are
second-order differential equations. However, the actual
time-space is almost without exception non-inertial, as it is
almost without exception that there exist (at least weak) fields,
waves, or forces perturbing an ideal inertial frame. Non-inertial
nature of the actual time-space is also supported by observations
of the practical astronomy that expansion of the reality occurs
with an acceleration. In other words, actually any real reference
frame is a non-inertial one; and such physical reality can be
described with a differential equation with time derivatives of
coordinates of the order exceeding two, which play the role of
additional variables. This is evidently beyond the scope of
Newtonian axiomatics. Aristotle's physics considered velocity to
be proportional to the applied force, hence the body dynamics was
described by first derivative differential equation. Newtonian
axiomaics postulates reference frames, where a free body maintains
the constant velocity of translational motion. In this case the
body dynamics is described with a second order differential
equation, with acceleration being proportional to force [1]. This
corresponds to the Lagrangian depending on coordinates and their
first derivatives (velocities) of the body, and Euler-Lagrange
equation resulting from the principle of the least action. This
model of the physical reality describes macrocosm fairly good, but
it fails to describe micro particles. Both Newtonian axiomatics
and the Second Law of Newton are invalid in microcosm. Only
averaged values of observable physical quantities yield in the
microcosm the approximate analog of the Second Law of Newton; this
is the so-called Ehrenfest's theorem. The Ehrenfest's equation
yields the averaged, rather than precise, ration between the
second time derivative of coordinate and the force, while to
describe the scatter of quantum observables the probability theory
apparatus is required. As the Newtonian dynamics is restricted to
the second order derivatives, while micro-objects must be
described with equations with additional variables, tending
Planck's constant to zero corresponds to neglecting these
variables. Hence, offering the model of extended Newtonian
dynamics, we consider classical and quantum theories with
additional variables, describing the body dynamics with higher
order differential equations. In our model the Lagrangian shall be
considered depending not only on coordinates and their first time
derivatives, but also on higher-order time derivatives of
coordinates. Classical dynamics of test particle motion with
higher-order time derivatives of coordinates was first described
in 1850 by M.Ostrogradskii [2] and is known as Ostrogradskii's
Canonical Formalism. Being a mathematician, M. Ostrogradskii
considered coordinate systems rather than reference frames. This
is just the case corresponding to a real reference frame
comprising both inertial and non-inertial reference frames. In a
general case, the Lagrangian takes on the form $(n\rightarrow
\infty)$

\begin{equation}
L=L(t,q,\dot{q},\ddot{q},...,\dot{q}^{(n)}).
\end{equation}

\section{Theory of Extended Newtonian Dynamics}

Let us consider in more detail this precise description of the
dynamics of body motion, taking into account of real reference
frames. To describe the extended dynamics of a body in an any
coordinate system (corresponding to arbitrary reference frame) let
us introduce concepts of kinematic state and kinematic invariant
of an arbitrary reference frame.

\textbf{Definition}: Kinematic state of a body is set by $n$-th time
derivative of coordinate. The kinematic state of the body is defined
provided the $n$-th time derivative of body coordinate is zero, the $(n-1)$%
-th time derivative of body coordinate being constant. In other
words, we consider the kinematic state of the body defined if
$(n-1)$-th time derivative of body coordinate is finite. Let us
note that a reference frame performing harmonic oscillations with
respect to an inertial reference frame does not possess any
definite kinematic state. Considering the dynamics of particles in
arbitrary reference frames, we suggest the following two
postulates.

\textbf{Postulate 1.} Kinematic state of a free body is invariable. This
means that if the $n$-th time derivative of a free body coordinate is zero,
the $(n-1)$-th time derivative of body coordinate is constant. That is,

\begin{equation}
\frac{d^{n}q}{dt^{n}}=0,\frac{d^{n-1}q}{dt^{n-1}}=const.
\end{equation}

In the extended model of dynamics, conversion from a reference
frame to another one will be defined as:
\begin{eqnarray}
q^{\prime } &=&q_{0}+\dot{q}t+\frac{1}{2!}\ddot{q}t^{2}+...+\frac{1}{n!}%
\dot {q}^{(n)}t^{n} \\
t^{\prime } &=&t.
\end{eqnarray}

\textbf{Postulate 2.} If the kinematic invariant of a reference frame is $n$%
-th time derivative of body coordinate, then the body dynamics is described
with the differential equation of the order $2n$:

\begin{equation}
\alpha _{2n}\dot {q}^{(2n)}+...+\alpha _{0}q=F(t,q,\dot{q},\ddot{q},...,%
\dot {q}^{(n)}).
\end{equation}

This means that the Lagrangian depends on $n$-th time derivative
of coordinate, so variation when applying the least action
principle will yield the order higher by a unity. Therefore, the
dynamics of a free body in a reference frame with $n$-th order
derivative being invariant shall be described with a differential
equation of the order $2n$. To consider dynamics of a body with an
observer in an arbitrary coordinate system (which corresponds to
the case of any reference frame), we apply the least action
principle, varying the action function for $n$-th order kinematic
invariant, we obtain the equation of the order $2n$:

\begin{equation}
\delta S=\delta \int L(t,\dot{q^{\prime }},q^{\prime })dt =\int
\sum_{n=0}^{N}(-1)^{n}\frac{d^{n}}{dt^{n}}\frac{\partial L}{\partial \dot{q}%
^{(n)}}\delta \dot{q}^{(n)}dt=0.
\end{equation}

Then the equation describing the dynamics of a body with $n$-invariant is a $%
2n$-order differential equation, and for the case of irreversible time arrow
we shall retain only even components. Expanding into Taylor's series the
function $q=q(t)$ yields:

\begin{equation}
q=q_{0}+\dot{q}t+\frac{1}{2!}\ddot{q}t^{2}+...+\frac{1}{n!}\dot{q}%
^{(n)}t^{n}.
\end{equation}

It is well known that the kinematic equation in inertial reference
frames of Newtonian physics contains the second time derivative of
coordinate, that is, acceleration:

\begin{equation}
q_{Newton}=q_{0}+vt+\frac{1}{2}at^{2}.
\end{equation}
Let us denote the additional terms with higher derivatives as

\begin{equation}
q_{r}=\frac{1}{3!}\dot{q}^{(3)}t^{3}+...+\frac{1}{n!}\dot{q}^{(n)}t^{n}.
\end{equation}

Then
\begin{equation}
q=q_{newton}+q_{r}.
\end{equation}

In our case, the discrepancy between descriptions of the two models is the
difference between the description of test particles in the model of
extended Newtonian dynamics with Lagrangian $L(t,q,\dot{q},\ddot{q},...,%
\dot {q}^{(n)},...)$ and Newtonian dynamics in inertial reference
frames with the Lagrangian $L(t,q,\dot{q})$:

\begin{equation}
\int [L(t,q,\dot{q},\ddot{q},...,\dot {q}^{(n)})-L(t,q,\dot{q})]dt=h,
\end{equation}
$h$ being the discrepancy (error) between descriptions by the two
models. Comparing this value with the uncertainty of measurement
in inertial reference frames, expressed by the Heisenberg
uncertainty relation, the equation (11) can be rewritten as
\begin{equation}
S(t,q,\dot{q},...\dot{q}^{(n)})-S(t,q,\dot{q})=h.
\end{equation}

In the classical mechanics, in inertial reference frames, the
Lagrangian depends only on the coordinates and their first time
derivatives. In the extended models, in real reference frames, the
Lagrangian depends not only on the coordinates and their first
time derivatives, but also on their higher derivatives. Applying
the least action principle [3], we obtain Euler-Lagrange equation
for the extended Newtonian dynamics model:

\begin{equation}
\sum_{n=0}^{N}(-1)^{n}\frac{d^{n}}{dt^{n}}\frac{\partial L}{\partial \dot{q}%
^{(n)}}=0,
\end{equation}

or
\begin{equation}
\frac{\partial L}{\partial q}-\frac{d}{dt}\frac{\partial L}{\partial \dot{q}}%
+\frac{d^{2}}{dt^{2}}\frac{\partial L}{\partial \ddot{q}}-...+(-1)^{N}\frac{%
d^{N}}{dt^{N}}\frac{\partial L}{\partial \dot{q}^{(N)}}=0.
\end{equation}
The Lagrangian will be expressed through quadratic functions of variables:
\begin{equation}
L=kq^{2}-k_{1}\dot{q}^{2}+k_{2}\ddot{q}^{2}-...+(-1)^{\alpha }k_{\alpha }%
\dot{q}^{(\alpha )2}=\sum_{\alpha =0}^{\infty }(-1)^{\alpha }k_{\alpha }\dot{%
q}^{(\alpha )2}.
\end{equation}
For our case, the action function will be:
\begin{equation}
S=q\frac{\partial L}{\partial q}-\dot{q}\frac{\partial L}{\partial \dot{q}}%
+...+(-1)^{\alpha }\dot{q}^{(\alpha )}\frac{\partial \dot{L}^{(\alpha )}}{%
\partial \dot{q}^{(\alpha )}}+...=\sum_{\alpha =0}^{\infty }(-1)^{\alpha }%
\dot{q}^{(\alpha )}\frac{d^{\alpha }}{dt^{\alpha }}\frac{\partial L}{%
\partial \dot{q}^{(\alpha )}}.
\end{equation}

Or
\begin{equation}
S=2kq^{2}-2k_{1}\dot{q}^{2}+2k_{2}\ddot{q}^{2}+...+2k_{\alpha }\dot{q}%
^{(\alpha )2}=2\sum_{\alpha =0}^{\infty }(-1)^{\alpha }k_{\alpha }\dot{q}%
^{(\alpha )2}.
\end{equation}

Introducing the notation

\begin{center}
\begin{equation}
F=\frac{\partial L}{\partial q},p=\frac{\partial L}{\partial \dot{q}}
\end{equation}%
\begin{equation}
F^{2}=\frac{\partial L}{\partial \ddot{q}},p^{3}=\frac{\partial L}{\partial
\dot{q}^{(3)}}
\end{equation}%
\begin{equation}
F^{4}=\frac{\partial L}{\partial \dot{q}^{(4)}},p^{5}=\frac{\partial L}{%
\partial \dot{q}^{(5)}}
\end{equation}%
.....\\[0pt]
\begin{equation}
F^{2n}=\frac{\partial L}{\partial \dot{q}^{(2n)}},p^{2n+1}=\frac{\partial L}{%
\partial \dot{q}^{(2n+1)}},
\end{equation}
\end{center}

we obtain the description of inertial forces for the extended Newtonian
dynamics model. The value of the resulting force accounting for inertial
forces can be expressed through momentums and their derivatives, expressing
the Second Law of Newton for the extended Newtonian dynamics model:
\begin{equation}
F-\frac{dp}{dt}+\frac{d^{2}}{dt^{2}}(F^{2}-\frac{dp^{3}}{dt})+\frac{d^{4}}{%
dt^{4}}(F^{4}-\frac{dp^{5}}{dt})+...\frac{d^{n}}{dt^{n}}(F^{n}-\frac{dp^{n+1}%
}{dt})=0.
\end{equation}

Expanding the force into Taylor series, we obtain:
\begin{equation}
F(t)=F_{0}+\dot{F}t+\frac{1}{2!}\ddot{F}t^{2}++...
\end{equation}

In other words, (22) can be written as
\begin{equation}
\sum_{n=0}^{\infty }\frac{d^{2n}}{dt^{2n}}(F^{2n}-\frac{d^{2n}p^{2n+1}}{%
dt^{2n}})=0.
\end{equation}
The action function takes on the form
\begin{equation}
S=\sum_{n=0}^{\infty }(-1)^{n}\dot{q}^{(n)}p^{n+1}=\sum_{n=0}^{N}(-1)^{n}%
\dot{q}^{(n)}\frac{\partial L}{\partial \dot{q}^{(n+1)}}.
\end{equation}

For this case, energy can be expressed as
\begin{equation}
E=\alpha _{0}q^{2}+\alpha _{1}\dot{q}^{2}+\alpha _{2}\ddot{q}^{2}+...+\alpha
_{n}\dot{q}^{(n)2}+...
\end{equation}%
Denoting the Appel's energy of acceleration [4] as $Q$, $\alpha _{n}$ being
constant factors, we obtain for kinetic energy and potential energy,
respectively,
\begin{eqnarray}
E &=&V+W+Q  \label{26} \\
V &=&\alpha _{0}q^{2}, \\
W &=&\alpha _{1}\dot{q}^{2} \\
Q &=&\alpha _{2}\ddot{q}^{2}+...+\alpha _{n}\dot{q}^{(n)2}+...
\end{eqnarray}%
The Hamilton-Jacobi equation for the action function will take on the form
\begin{equation}
-\frac{\partial S}{\partial t}=\frac{(\nabla S)^{2}}{2m}+V+Q,
\end{equation}%
The first addend in (30) is the so-called Appel's energy of
acceleration [4]. Let us compare $Q$ with the quantum potential
[5] and complement the equation (31) with the continuity equation.
If $Q\approx \alpha _{2}\frac{\nabla ^{2}S}{m^{2}}$ (here, the
value of the constant is chosen $\alpha _{2}=\frac{i\hbar m}{2}$).
Hence, in the first approximation we obtain for the function
\begin{equation}
\psi =e^{\frac{i}{\hbar }S},
\end{equation}%
the Schroedinger equation\bigskip
\begin{equation}
i\hbar \frac{\partial \psi }{\partial t}=\frac{\hbar ^{2}}{2m}\nabla
^{2}\psi +V\psi .
\end{equation}

\section{Conclusions}

Our case corresponds to Lagrangian $L(t,q,\dot{q},\ddot{q},...,\dot {q}%
^{(n)},...)$, depending on coordinates, velocities and higher time
derivatives, which we call additional variables, extra addends, or
hidden variables. In arbitrary reference frames (including
non-inertial ones) additional variables (addends) appear in the
form of higher time derivatives of coordinates, which complement
both classical and quantum physics. We call these additional
addends, or variables, constituting the higher time derivatives of
coordinates, addition variables or hidden variables, complementing
the description of particles. It should be noted that these hidden
variables can be used to complement the quantum description
without violating von Neumann theorem, as this theorem is not
applied for non-linear reference frames, while the extended
Newtonian dynamics model assumes employing any reference frames,
including non-linear ones. Comparing the generalized
Hamilton-Jacobi equation
\begin{equation}
\-\frac{\partial S}{\partial t}=\frac{(\nabla S)^{2}}{2m}+V+Q,
\end{equation}%
$Q$ being the additional variables with higher derivatives, with
the quantum Bohm's potential, one can conclude that neglecting
higher-order time derivatives of coordinates brings about
incompleteness of physical reality description. The coordinate
derivative of Q determines the quantum force. This means that
complete description of physical reality requires considering
differential equations of the order exceeding second; uncertainty
of the position of the particle under investigation shall be
attributed to fluctuations of the reference body and reference
frame associated with it. Hence, the differential equation
describing this case shall be of the order exceeding second. In
this case, uncertainty of a micro objects description is follow by
incompleteness of the description of the physical reality by
Newtonian physics, that is, the lack of a complete description
with additional variables in the form of higher time derivatives
of coordinates. The contemporary physics presupposes employment of
predominantly inertial reference frames; however, such a frame is
very hard to obtain, as there always exist external perturbative
effects, for example, gravitational forces, fields, or waves. In
this case, the relativity principle enables transfer from the
gravitational forces or waves to inertial forces. For example, if
we consider a spaceship with two observers in different cabins,
one can see that this system is non-ideal, the inertial forces (or
pseudo-forces) could constitute additional variables here. In this
case, superposition of the two distributions obtained by the
observers could yield a non-zero correlation factor, though each
of the two observations has a seemingly random nature. If the fact
that the reference frame is non-inertial and hence there exist
additional variables in the form of inertial effects is ignored,
then non-local correlation of seemingly independent observations
would seem surprising. This example could visualize not only the
interference of corpuscle particles, but also the non-local
character of quantum correlations when considering the effects of
entanglement.

\end{document}